\documentclass[10pt,twocolumn,aps,prd,amsmath,amssymb,oneside,showpacs,nofootinbib,preprintnumbers]{revtex4}
\usepackage[T1]{fontenc}

\newcommand{\im}{\mathrm{i}}  

\newcommand{\xd}{\mathrm{d}}
\newcommand{\xD}{\mathcal{D}} 

\newcommand{\be}{\begin{equation}}
\newcommand{\ee}{\end{equation}}
\newcommand{\bea}{\begin{eqnarray}}
\newcommand{\eea}{\end{eqnarray}}

\begin{document}

\title{S-matrix in de Sitter spacetime from general boundary quantum field theory}
\author{Daniele Colosi}\email{colosi@matmor.unam.mx}
\affiliation{Instituto de Matem\'aticas, UNAM, Campus Morelia, C.P.~58190, Morelia, Michoac\'an, Mexico}
\date{\today}
\preprint{UNAM-IM-MOR-2009-2}

\begin{abstract}
A new quantization scheme for a massive scalar field in de Sitter spacetime is proposed, based on the general boundary formulation of quantum field theory. We show that the general interacting theory can be consistently described in terms of the $S$-matrix for spatial asymptotic states. The new $S$-matrix results to be equivalent to the standard one in situations where both apply. This is due to the existence of an isomorphism between the corresponding asymptotic state spaces.
\end{abstract}

\maketitle

The most studied example of quantum field theory in curved spacetime is probably the theory of a scalar field in de Sitter space. Indeed the model is simple enough to be solved analytically and therefore the properties of the field can be studied in detail. In particular the thermal properties of the vacuum state, related to the phenomenon of particle creation, have been considered \cite{Gibbons:1977mu}. Moreover, the quantum dynamics of the field is of special importance for inflationary cosmological models where de Sitter space describes a universe in exponential expansion \cite{Mukhanov:2005sc}. De Sitter space has also attracted new interest in connection with the conjecture of the dS/CFT correspondence proposed almost a decade ago by Strominger \cite{Strominger:2001pn}.

In this letter we expose a new quantization scheme for a massive scalar field in de Sitter spacetime based on the general boundary formulation (GBF) of quantum field theory. The results presented here are mainly intended as a contribution to the GBF program, and represent indeed the first study of QFT in a curved spacetime within the GBF. Furthermore, the quantization scheme we introduce provides a new framework to analyze the results obtained or proposed so far in literature.

In a series of papers \cite{Oe:timelike,Oe:GBQFT,Oe:KGtl,CoOe:spsmatrix,CoOe:smatrixgbf,CoOe:smatrix2d} it has been shown that the GBF
 provides a viable description of the dynamics of quantized fields. Not only this new formulation can recover known results of standard QFT, but more interesting the GBF can handle situations where the methods of standard QFT fail. 
In QFT, dynamics is described in terms of the evolution of initial data from an initial Cauchy surface to a final Cauchy surface. Therefore, this standard picture
involves a spacetime region bounded by the disjoint union of two 
Cauchy surfaces. In the GBF
evolution acquires a more general description: The boundary of the spacetime region where dynamics takes place can have arbitrary form and is not required to reduce to the disjoint union of two Cauchy surfaces.

The main novelty of the GBF resides in associating Hilbert spaces of states to arbitrary hypersurfaces in spacetime and amplitudes to spacetime regions and states living on their boundaries. For a region $M$ of spacetime, the amplitude is a map from the Hilbert space ${\cal H}_{\partial M}$ associated with the boundary $\partial M$ of the region to the complex numbers. The formal expression of the amplitude $\rho$, for a state $\psi \in {\cal H}_{\partial M}$, is given in terms of the Feynman path integral combined with the Schr\"odinger representation for quantum states,
\be
\rho_{M}(\psi)= \int \xD \varphi \, \psi(\varphi) \int_{\phi|_{\partial M}=\varphi} \xD\phi\, e^{\im S_{M}(\phi)},
\label{eq:rho}
\ee
where the outer integral is over all field configurations $\varphi$ on $\partial M$, and the inner integral is over all field configurations $\phi$ in the spacetime region $M$ matching $\varphi$ on the boundary.
Finally, a physical interpretation can be given to such amplitudes and an appropriate notion of probability can be extracted from them \cite{Oe:GBQFT,Oe:KGtl}.

So far this formalism has been applied only to the study of flat-spacetime-based QFT. There the standard $S$-matrix for an interacting scalar field has been shown to be equivalent to the one derived for free asymptotic quantum states at spatial infinity. The notion of spatial asymptotic state comes from the geometry considered: In particular, in Minkowski spacetime states were defined on an hypercylinder, namely the boundary of a threeball in space extended over all of time, and then the radius of the ball was sent to infinity. 

The structure of this work follows that of \cite{CoOe:smatrixgbf}. So, in the following we will evaluate the $S$-matrix for coherent states on spacelike hypersurfaces of constant conformal de Sitter time. Then, the asymptotic amplitude will be derived for coherent states defined on an analogue of the hypercylinder in de Sitter space. Finally, by constructing an isomorphism between the respective state spaces, we prove the equivalence of these two types of amplitude.


We consider de Sitter spacetime with the metric
\be
\xd s^2 = \frac{1}{H^2t^2} \left( \xd t^2 - \xd \underline{x}^2  \right),
\label{eq:dSmetric}
\ee
where $H$ is the Hubble constant, the conformal time $t$ takes values in the interval $]0,\infty[$ and $\underline{x} \in {\mathbb R}^3$ are coordinates on the equal time hypersurfaces. Such coordinates cover only half of de Sitter spacetime. The remaining half can be included by simply extending the domain of $t$ to $]-\infty,\infty[$, \cite{BiDa:qfcs}. We start with the derivation of the standard transition amplitude by computing the amplitude (\ref{eq:rho}) for a spacetime region $M$ bounded by two equal time hypersurfaces, $\Sigma_1$ and $\Sigma_2$, defined respectively by the values $t_1$ and $t_2$ of the conformal time $t$: $M = [t_1,t_2] \times \mathbb{R}^3$. Then the state space associated with the boundary $\partial M = \Sigma_1 \cup \Sigma_2$ results to be the tensor product ${\cal H}_1 \otimes {\cal H}_2^*$ of the Hilbert spaces defined on the hypersurfaces $\Sigma_1$ and $\Sigma_2$. Following \cite{CoOe:smatrixgbf} we introduce coherent states; their wave function at time t is parametrized by a complex function $\xi$ on momentum space, and in the interaction picture their form is
\begin{multline}
\psi_{t,\xi} (\varphi) =  K_{t, \xi} \\ 
\times \exp \left( \int \frac{\xd ^3 x \, \xd^3 k}{(2 \pi)^3} \, \xi(\underline{k}) \,  \frac{e^{\im \underline{k} \cdot \underline{x}}}{H_{\nu}(k t) t^{3/2}} \,  \varphi(\underline{x}) \right) 
\psi_{t,0}(\varphi),
\end{multline}
where $K_{t, \xi}$ is a normalization factor, $\psi_{t,0}(\varphi)$ is the vacuum wave function derived in \cite{Co:vacuum}, $H_{\nu}$ is the Hankel function of order $\nu = \sqrt{\frac{9}{4} - \frac{M^2}{H^2}}$, $M$ denotes the mass of the scalar field and we assume $\nu$ real.

Consider first the free theory. The amplitude (\ref{eq:rho}), denoted by the subscript $0$, associated with the tensor product of coherent states $\psi_{t_1,\xi_1} \otimes \overline{\psi_{t_2,\xi_2}}$ is independent of times $t_1$ and $t_2$ and can be written as
\begin{multline}
\rho_{M,0}(\psi_{\xi_1} \otimes \overline{\psi_{ \xi_2}}) = 
\exp \left(  \frac{\pi H^2}{4} \int \frac{\xd^3 k}{(2 \pi)^3} \,  \right. \\
\times \left. \left(  \overline{\xi_{2}(\underline{k})} \xi_{1}(\underline{k}) - \frac{1}{2}|\xi_{1}(\underline{k})|^2 - \frac{1}{2} |\xi_{2}(\underline{k})|^2  \right) \right).
\label{eq:freeampl}
\end{multline}
Because of independence of initial and final time, the above expression represents the $S$-matrix describing the transition from the coherent state defined by $\xi_1$ (in the asymptotic past) to the coherent state defined by $\xi_2$ (in the asymptotic future).

As an intermediate step in the computation of the $S$-matrix for the general interacting theory, we consider the interaction of the scalar field with a source field $\mu$ of the form $\int \xd^4 x \sqrt{-g} \, \phi(x) \, \mu(x)$, and we assume that $\mu$ vanishes outside the spacetime region considered here, namely $\mu |_{t \notin ]t_1,t_2[} = 0$. Adding such interaction term to the free action yields for the amplitude (\ref{eq:rho}), denoted by the subscript $\mu$, the result
\begin{multline}
\rho_{M,\mu}(\psi_{\xi_1} \otimes \overline{\psi_{\xi_2}}) = \\
\rho_{M,0}(\psi_{ \xi_1} \otimes \overline{\psi_{\xi_2}})  \exp \left(  \int \xd ^4 x \sqrt{-g}  \mu(x) \hat{\xi}(x) \right)   \\ 
\times \exp \left(\frac{\im}{2}  \int \xd ^4 x  \sqrt{-g}  \, \mu(x) \, \gamma(x) \right),
\label{eq:srcampl} 
\end{multline}
where $g$ is the determinant of the metric (\ref{eq:dSmetric}), $\hat{\xi}$ is the complex solution of the Klein-Gordon equation determined by the initial and final coherent states,
\begin{multline}
\hat{\xi}(x) = \frac{\im \pi H^2}{4} \int \frac{ \xd^3 k}{(2 \pi)^3} \left(\xi_1(\underline{k}) \, e^{\im \underline{k} \cdot \underline{x}} \, t^{3/2} \,  \overline{H_{\nu}(k t)} \right. \\
\left. + \, \overline{\xi_2(\underline{k})} \, e^{-\im \underline{k} \cdot \underline{x}}\,  t^{3/2} \,  H_{\nu}(k t) \right).
\label{eq:hatxi}
\end{multline}
The function $\gamma$ in the last exponential of (\ref{eq:srcampl}) is the solution of the inhomogeneous Klein-Gordon equation,
\be
(\Box + M^2) \gamma(x) = \mu(x),
\label{eq:inKG}
\ee
with the following boundary conditions,
\begin{multline}
\gamma(t, \underline{x}) |_{t<t_1} = t^{3/2} \, H_{\nu}(k t) \, \frac{\im \pi H^2}{4  }  \\
\times \int_{t_1}^{t_2} \xd t' \sqrt{-g'}   (t')^{3/2}  \overline{H_{\nu}(k t')}  \mu(t',\underline{x}), \label{eq:Fbc1}
\end{multline}
for early times $t$ before the source is switched on, and
\begin{multline}
\gamma(t, \underline{x}) |_{t>t_2} = t^{3/2} \, \overline{H_{\nu}(k t)} \, \frac{\im \pi H^2}{4  } \\
\times \int_{t_1}^{t_2} \xd t'  \sqrt{-g'}   (t')^{3/2}  H_{\nu}(k t')  \mu(t',\underline{x}),\label{eq:Fbc2}
\end{multline}
for late times $t$ after the source is switched off. In the above expressions $g'$ is the determinant of the metric (\ref{eq:dSmetric}) expressed in the coordinates $(t', \underline{x})$, and the Bessel functions are to be understood as operators via the mode decomposition of the source field. It is convenient the write $\gamma$ in the form
\be
\gamma(x) = \int \xd^4 x' \sqrt{-g'} \, G(x,x') \, \mu(x'),
\label{eq:gamma}
\ee
and $G$ is the Green function, solution of the equation $(\Box + M^2)G(x,x') = (-g)^{-1/2} \delta^4(x-x')$, given by
\begin{multline}
G(x,x')= \frac{H^2}{ 16 \pi} \left( \frac{1}{4} - \nu^2 \right) \sec(\nu \pi)\\
\times  F \left(  \frac{3}{2}-\nu ,  \frac{3}{2} + \nu ; 2 ; \frac{1+p-  \im 0}{2}\right),
\label{eq:propagator}
\end{multline}
where $F$ is the hypergeometric function and $p= \frac{t^2+t'^2-|\underline{x}- \underline{x}'|^2}{2 t't}$. The above expression coincides with the expression of the Feynman propagator in de Sitter space computed in \cite{Schomblond:1976xc,Bunch:1978yq}, and we can therefore interpret the conditions (\ref{eq:Fbc1},\ref{eq:Fbc2}) has the Feynman boundary conditions. The form (\ref{eq:srcampl}) of the $S$-matrix in the presence of a source field is similar to the one obtained by the path integral in the holomorphic representation \cite{FaSl:gaugeqft}.

Finally we use functional methods to express a general interaction in terms of the source interaction,
\begin{multline}
\int \xd^4 x \, \sqrt{-g} \, V(x, \phi(x)) = \\
\int \xd^4 x \, V \left( x, \frac{\partial}{\partial \mu(x)}\right) \int \xd^4 y \sqrt{-g} \, \phi(y) \mu(y) \bigg|_{\mu =0}.
\label{eq:genint}
\end{multline}
Assuming that the interaction vanishes for $t$ outside the interval $]t_1,t_2[$, the amplitude (\ref{eq:rho}), now indicated with the subscript $V$, takes the form
\begin{multline}
\rho_{M,V}(\psi_{\xi_1} \otimes \overline{\psi_{\xi_2}}) = \exp \left( \im \int \xd^4 x \, V \left( x, - \im \frac{\partial}{\partial \mu(x)}\right) \right) \\
\times \rho_{M, \mu}(\psi_{\xi_1} \otimes \overline{\psi_{\xi_2}})\bigg|_{\mu =0}.
\label{eq:genampl}
\end{multline}
This expression is independent of $t_1$ and $t_2$ and consequently the restriction on $V$ introduced above can be removed. Moreover, the limit of asymptotic times is trivial and (\ref{eq:genampl}) is then interpreted as the $S$-matrix for the general interacting theory.

The second geometry we are interested in is conveniently described in terms of spherical coordinates, in which the metric (\ref{eq:dSmetric}) takes the form
\be
\xd s^2 = \frac{1}{H^2 t^2} \left( \xd t^2 - \xd r^2 - r^2 \xd \Omega^2 \right),
\label{eq:dSmetric2}
\ee
where $\xd \Omega^2$ is the metric on a unit sphere.
We will now compute the amplitude (\ref{eq:rho}) associated with the spacetime region $M$ bounded by the hypersurface of constant radius, $r =R$. Hence, $M$ has one connected boundary that we call the \textit{hypercylinder} in analogy with the notion of the hypercylinder introduced in \cite{Oe:KGtl}. We proceed as before by considering coherent states defined in the Hilbert space ${\cal H}_R$ associated with the hypercylinder, and with wave function in the interaction picture given by
\begin{multline}
\psi_{R,\eta} (\varphi) = K_{R, \eta} \exp \left( \int \xd t \, \xd \Omega \, \xd k \sum_{l,m} \right. \\
\left. \times \eta_{l,m}(k)\frac{t^{-1/2} Z_{\nu}(k t) Y_l^{-m}(\Omega)}{h_l(k R)} \varphi(t, \Omega) \right) \psi_{R,0}(\varphi),
\end{multline}
where $\eta$ is the complex function on momentum space that parametrizes the coherent state, $\psi_{R,0}$ the vacuum wave function on the hypercylinder of radius $R$ and $K_{R, \eta}$ a normalization factor. Here $Z_{\nu}$ denotes the Bessel function of the first or second kind, $Y_l^m$ the spherical harmonic and $h_l$ the spherical Bessel function of the third kind. 

The free amplitude for such state reads,
\begin{multline}
\rho_{M,0}(\psi_{ \eta} ) =   \exp \left( - \frac{H^2}{4} \int \xd k \sum_{l,m}  k^2  \right. \\
 \times \left[ |\eta_{l,m}(k)|^2  - \eta_{l,m}(k) \, \eta_{l,-m}(k)  \right]  \Bigg),
\label{eq:freeamplhyp}
\end{multline}
and is independent of the radius $R$. 

As before, we now look at the interaction with a source field $\mu$. Requiring this field to be confined in the interior of the hypercylinder, the amplitude for the coherent state $\psi_{\eta}$ turns out to be
\begin{multline}
\rho_{M,\mu}(\psi_{ \eta} ) = 
\rho_{M,0}(\psi_{ \eta} )   \exp \left(  \int \xd ^4 x \sqrt{-g}  \mu(x) \hat{\eta}(x) \right)  \\ 
\times \exp \left(\frac{\im}{2}  \int \xd ^4 x  \sqrt{-g}  \, \gamma(x) \, \mu(x) \right).
\label{eq:srcamplhyp} 
\end{multline}
$\hat{\eta}$ is the complex solution of the Klein-Gordon equation given by
\begin{multline}
\hat{\eta}(x) = \im H^2 \int \xd k \, k \sum_{l,m} \, t^{3/2} Z_{\nu}(k t) \\
\times Y_l^m(\Omega) \,  j_l(kr)  \, \eta_{l,m}(k),
\label{eq:hateta}
\end{multline}
where $j_l$ is the spherical Bessel function of the first kind. 
The function $\gamma$ in the last line of (\ref{eq:srcamplhyp}) satisfies the inhomogeneous Klein-Gordon equation (\ref{eq:inKG}), and can therefore be written via the Green function as in (\ref{eq:gamma}).
The Green function $G$, defined on the hypercylinder, turns out to be the same Green function that appears in (\ref{eq:gamma}), namely the Feynman propagator (\ref{eq:propagator}). The boundary condition satisfied by $\gamma$ can then be interpreted as the Feynman boundary condition on the hypercylinder, valid for large radius $r$ outside the source field,
\begin{multline}
\gamma(t,r, \Omega)\big|_{r>R} = k \,\im \, h_l(k r) \\
\times \int_0^R \xd r' \, (r')^2 \sqrt{-g'} \,t^2 H^2 \, j_l(k r') \mu(t,r', \Omega).
\end{multline}
$g'$ denotes the determinant of the metric (\ref{eq:dSmetric2}) in the coordinates $(t,r', \Omega)$, and the Bessel functions are to be understood as operators acting on the mode expansion of $\mu$. 
Again, we notice that no dependence on the radius $R$ is present in the amplitude (\ref{eq:srcamplhyp}). 

To conclude, we apply the same functional techniques as before, expressing the general interaction as in (\ref{eq:genint}). Assuming that the interaction now vanishes outside the hypercylinder, we can write the amplitude for the general interacting theory as
\begin{multline}
\rho_{M,V}(\psi_{\eta}) = \exp \left( \im \int \xd^4 x \, V \left( x, - \im \frac{\partial}{\partial \mu(x)}\right) \right) \\
\times \rho_{M, \mu}(\psi_{\eta})\bigg|_{\mu =0}.
\label{eq:genamplhyp}
\end{multline}
Since $R$ does not appear, the cutoff on the interaction can be dropped. Being the limit $R \rightarrow \infty$ trivial, we interpret (\ref{eq:genamplhyp}) as the asymptotic amplitude of the general interacting theory for the coherent state $\psi_{\eta}$.

Having computed the asymptotic amplitudes in the two geometries considered here, we now want to analyze their relation. To this aim we adopt an approach analogue to that used in \cite{CoOe:smatrixgbf}. We focus our attention on the expression of the amplitudes for the source interaction in both settings, i.e. (\ref{eq:srcampl}) and (\ref{eq:srcamplhyp}). Considering the same source field in the two cases, namely a source bounded in space and in time, we notice that the last terms of the amplitudes coincide because in the functions $\gamma$ the same propagator (\ref{eq:propagator}) appears. We turn to the second second terms in (\ref{eq:srcampl}) and (\ref{eq:srcamplhyp}): They coincide if and only if the complex solutions to Klein-Gordon equation, $\hat{\xi}$ and $\hat{\eta}$, coincide. It turns out that this equality, namely $\hat{\xi} = \hat{\eta}$, defines an isomorphic map between the state spaces of the two theories, i.e. the Hilbert space ${\cal H}_1 \otimes {\cal H}_2^*$ associated with the boundary of the spacetime region $M = [t_1,t_2] \times \mathbb{R}^3$ and the Hilbert space ${\cal H}_R$ associated with the hypercylinder. Hence, under the isomorphism we have: $\psi_{\xi_1} \otimes \overline{\psi_{\xi_2}} \cong \psi_{\eta}$. We are left with the first term appearing in (\ref{eq:srcampl}) and (\ref{eq:srcamplhyp}), the free amplitudes in the two settings given by (\ref{eq:freeampl}) and (\ref{eq:freeamplhyp}). It is not difficult to show that these free amplitudes are equal under the isomorphism. For example, expressing (\ref{eq:freeamplhyp}) in terms of the function $\hat{\eta}$, we substitute $\hat{\eta}$ with the expression (\ref{eq:hatxi}) of $\hat{\xi}$ and obtain (\ref{eq:freeampl}):
\be
\rho_{M,0}(\psi_{\eta})\big|_{\hat{\eta} = \hat{\xi}} = \rho_{M,0}(\psi_{\xi_1} \otimes \overline{\psi_{ \xi_2}}).
\ee
We can then conclude the equivalence of the asymptotic amplitudes, interpreted as $S$-matrices, for the general interacting theory, under the isomorphism. Such equivalence offers the possibility to study scattering processes in de Sitter space from a new perspective. Indeed the amplitude for a transition from an in-state with $m$ particles to an out-state with $n$ particles can be mapped to the amplitude for an $(m+n)$-particle state defined on the hypercylinder, and the physical probabilities extracted from the $S$-matrices of the two descriptions are the same. We recover here results analogous to those previously obtained in Minkowski \cite{CoOe:smatrixgbf} and Euclidean spacetime \cite{CoOe:smatrix2d}, and the conclusions discussed there can be exported, mutatis mutandi, to de Sitter space.




\begin{acknowledgments}

I am grateful to Robert Oeckl for helpful discussions and comments on an earlier draft of this letter. This work was supported in part by CONACyT grant 49093.

\end{acknowledgments}

\bibliographystyle{amsordx}
\bibliography{stdrefs,refs2}

\end{document}